# Parallel plate capacitor analogy of equatorial plasma bubble and associated fringe fields with implications to equatorial valley region irregularities

S. Mukherjee[1] and A. K. Patra[2]

[1]Department of Physics, Indian Institute of Technology, Kharagpur, India

[2]National Atmospheric Research Laboratory, Gadanki, India

Corresponding Author: A K Patra

National Atmospheric Research Laboratory,

Gadanki 517112, AP, India

Phone: +91-8585-272010

Email: akpatra@narl.gov.in

**Abstract.**

VHF radar echoes from the valley region plasma irregularities, displaying ascending pattern, are often observed during the active phase of equatorial plasma bubble at the magnetic equator and have been attributed to bubble related fringe field effect. These irregularities, however, are not observed at a few degrees away from the equator. In this paper, we attempt to understand this contrasting observational result by comparing fringe field at the equator and low latitudes. We use parallel plate capacitor analogy of equatorial plasma bubble and choose a few capacitor configurations, consistent with commonly observed dimension and magnetic field-aligned property of plasma bubble, for computing fringe field. Results show that while fringe field decreases significantly with decreasing altitude as expected, remarkably it decreases with latitude, which clearly indicates the role of magnetic field-aligned property of plasma bubble in reducing the magnitude of fringe field at low latitudes compared to that of the equator. The results are presented and discussed in the light of plasma bubble associated fringe field induced plasma irregularities in the valley region to account for the observed difference at the equator and low latitudes.

## 1. Introduction

Radar echoes from the valley region, displaying ascending echoing pattern, in association with F region plume structures in the Jicamarca radar observations [e.g., Woodman and LaHoz, 1976; Hysell et al., 1990; Woodman and Chau, 2001], continue to be illusive in terms of their origin. For illustration, an example of Jicamarca observation, presented earlier by Woodman and Chau [2001], is reproduced in Figure 1. Note the ascending pattern in the valley region echoing regions. Valley region echoes of similar kind were also reported from Sao Luis, a Brazilian equatorial location [Alam Kherani et al., 2012]. These observations imply the presence of small-scale plasma irregularities with scale sizes (3-5 m) that match the Bragg scattering condition ($\lambda_{irregularity} = \lambda_{radar}/2$) in the valley region. These meter-scale irregularities are believed to be generated through a turbulence cascade from irregularities at larger scale than a few meters. Woodman [1993] suggested that these valley region irregularities could be generated while E-region plasma is up-lifted by the fringe electric fields associated with the growth phase of the equatorial plasma bubble [Zalesak and Ossakow, 1980]. Although this interpretation was quite consistent with the facts that plasma inside the depleted regions has extremely low density and comprises of metallic ions [Hanson and Sanatani, 1971; Hanson et al., 1972], which are formed in the E region by meteoric ablation process, no analysis on the fringe field and their effect were made at that time. Later, Alam Kherani et al. [2002, 2004], using numerical simulation, attempted to study the effect of fringe fields associated with the generalized Rayleigh-Taylor instability on the valley and E regions in order to explain the valley region irregularities. Alam Kherani et al. [2004], however, found that the fringe fields could not penetrate into the E region due to increasing collision with decreasing altitude, but could reasonably penetrate to the upper E region (120 km). They suggested that the E region irregularities occurring close to 120 km can be effectively pulled up by the fringe field to the valley region and above explaining their occurrence in the valley region.

Valley region echoing structures of the type observed at Jicamarca and Sao Luis [e.g., Woodman and Chau, 2001; Alam Kherani et al., 2012], however, have not been observed at Gadanki, located at 6.5$^{o}$ N magnetic latitude [Patra et al., 2007]. An example of Gadanki observations is shown in Figure 2. Note that no valley region echoes were observed despite the fact that fully blown F region plume structures were observed in the post-sunset hours.

Given the fact that plasma bubble grows as flux tube integrated entity, i.e., it grows vertically and along the magnetic field simultaneously, not observing these irregularities at off-equatorial location raises an important question on the validity of the interpretation of the equatorial ascending type valley region irregularities in terms of plasma bubble related fringe field. Fringe field computations done by Alam Kherani et al. [2002, 2004], however, have been performed for the equatorial magnetic field geometry and it is not known how the fringe fields would vary from horizontal- to inclined- magnetic field geometry.

In this work, we use a parallel plate capacitor analogy of the equatorial plasma bubble and estimate fringe fields as a function of decreasing height, following analytical expressions given by Parker [2002]. The shapes of the capacitor plates are chosen in such a manner that the penetration of fringe fields can be defined for both horizontal and inclined magnetic field geometries for making a comparison between the two. Results show noticeable reduction in the fringe field as a function of latitude and decreasing altitude. These results are presented and discussed in the light of current understanding of the fringe field effects in the ionospheric plasma transport and formation/upward transport of irregularities.

## 2. Fringe field in parallel plate capacitor

When a parallel plate capacitor with a plate area of A and an inter-plate separation distance d is filled with a dielectric of permittivity ε and subjected to a potential V, the capacitance and electric field between the plates can be calculated as:

$$C = Q/V \qquad (1)$$

$$E = V/d = Q/\varepsilon A, \qquad (2)$$

where Q is the total charge accumulated on the plate. Measured capacitance and electric field, however, differ from theoretical value. These effects are explained in terms of fringe field effects.

Parker [2002] provided a theoretical framework and obtained analytic expressions for electric field outside a strip capacitor. He used a parallel plate capacitor comprising of two plates, each having length L and width W (W<<L) and plate separation of d (d<<L and W), which are assigned with potentials of ±V/2. Parker [2002] simplified the above mentioned configuration in such a way that the configuration becomes two-dimensional, i.e., cross section of the parallel plates is in the x=0 plane, d parallel to z-axis and L parallel to y-axis. Then by using the symmetry about zero potential, which is along the z-axis, Parker [2002] obtained two components of electric field as:

$$E_z(y,z) = \left(\frac{V}{2\pi}\right)\left[\frac{\frac{W}{2}+y}{\left[z^2+\left(\frac{W}{2}+y\right)^2\right]}+\frac{\frac{W}{2}+y}{\left[z^2+\left(\frac{W}{2}-y\right)^2\right]}\right], \qquad (3)$$

$$E_y(y,z) = \left(\frac{-Vz}{2\pi}\right)\left[\frac{1}{\left[z^2+\left(\frac{W}{2}+y\right)^2\right]}-\frac{1}{\left[z^2+\left(\frac{W}{2}-y\right)^2\right]}\right], \qquad (4)$$

where $E_z$ and $E_y$ are the electric field components along d and L, respectively. The fields become infinite at the ends because infinitely thin plates are assumed. Electric field lines outside the plate in the y-z plane are given as:

$$\frac{dz}{dy} = \frac{E_z}{E_y} = \frac{z^2 - y^2 + \frac{W^2}{4}}{2yz}, \qquad (5)$$

Then by shifting the origin along y-axis by W/2, i.e., $z' = z,\ y' = y - \frac{W}{2}$, equation 5 for arbitrary value of W is rewritten as:

$$\frac{dz'}{dy'} = \frac{E_z}{E_y} = \frac{z'^2 - y'^2 - y'W}{2y'z' + z'W} \tag{6}$$

The solution of equation 6, for z' > 0 is given as [Parker, 2002]:

$$z'(y') = \left(\frac{1}{2}\right)\sqrt{W + 2y'}\sqrt{C - \frac{W^2 + 2Wy' + 4y'^2}{W + 2y'}}, \tag{7}$$

In the case of C > W, the allowed range of $y'$ for lines in the right half of the y-z plane i.e., $y'_- < y' < y'_+$, is given as :

$$y'_{\mp} = \frac{C - W}{4}\left[1 \mp \sqrt{1 + \frac{4W}{C - W}}\right], \tag{8}$$

More details on the formulation and mathematical treatment used in deriving fringe fields can be found in Parker [2002].

**3. Fringe field computation**

In the present work, we have performed fringe field computation for two different configurations of parallel plate capacitor. These two configurations are shown in Figures 3a and 3b, respectively. Configuration-1 consists of two rectangular parallel plates having length (L), breadth (W) and inter-plate separation (d)) in such a way that they are equivalent to 400 km in the vertical, 1200 km in the north-south and 100 km in the east-west directions, respectively. Configuration-2 consists of two plates with similar dimensions as that of configuration-1 but with the bottom edge having a curvature, which is an approximation for low latitude magnetic field geometry. The curved edge is chosen in such a way that it is elongated along north-south and symmetric about the magnetic equator. Considering the facts that plasma bubble maps along magnetic field lines and magnetic field lines at low latitudes have finite dip angle, this configuration has been chosen. Thus while configuration-1 is applicable to the magnetic equator, where the magnetic field is horizontal, configuration-2 can be used for low latitudes, where magnetic field lines are tilted, as well as at the equator. Moreover, fringe fields of these configurations can also be used to understand the equatorial and low latitude observations of valley region irregularities mentioned in the introduction.

In the adopted coordinate system, described in section 2, x is parallel to north-south, y is parallel to vertically downward direction and z is parallel to the inter-plate separation, which is parallel to east-west direction. In the present case, we are primarily concerned with the east-west electric field, $E_z$ that is responsible for the up-down motion of plasma. $E_z$ component of fringe field is computed following equation 3 as a function of decreasing altitude, which is parallel to y axis and in the vertical downward direction. The coordinates through which the field lines would pass are obtained from equations 7 and 8.

Equations 3-8, however, are applicable for a strip capacitor and here we need to obtain the fringe fields for configuration-1 and configuration-2, which have finite W (in the present case 1200 km). One possible way is to integrate the fringe fields of several such capacitors. In the present study, we have considered seven plate capacitors for simplicity and integrated the east-west components of electric fields (i.e., $E_z$) outside the plate capacitor in the y-z plane (i.e., east-west component of fringe field). For the second configuration seven small plate capacitors have been used to obtain an equivalent plate capacitor. These plates are oriented in such a manner that they approximately form the shape of the configuration-2. Then the resultant fringe fields are computed.

The capacitors shown in Figure 3 are assumed to be located at an altitude of 300 km so that fringe fields below this altitude can be computed for their application to the ionospheric phenomenon chosen in this work. Given the fact the base of plasma bubble lies around 300 km, the above assumption is quite realistic.

**4. Results and discussion**

First, we address how fringe field varies along north-south for configuration-1 and configuration-2. Figures 4(a) and (b) show peak values of fringe field, which peak at mid-points of the capacitors, for configuration-1 and configuration-2, respectively. Fringe fields have been computed for L=400 km, W=1200 km, and d=100 km. All fringe fields have been normalized to unity and plotted in different scales for different heights for clarity. In case of configuration-1, fringe field decreases with altitude, but does not change along north-south. In contrast to this, fringe field in configuration-2 not only decreases with height, it decreases remarkably along north-south. Fringe field attains a maximum value at the centre i.e., at the equator and the percentage decrease in the normalized electric field over the equator at altitudes of 200 km, 150 km, and 100 km with respect to that at 250 km are 86.73%, 94.83% and 97.03%, respectively. In case of configuration-2, fringe field is found to decrease along north-south also. For example, corresponding to 250 km altitude, the fringe field at a distance of 1200 km from the equator becomes 57% of that of the equator. At a similar distance from the equator, fringe field reduces remarkably at lower altitudes and at an altitude of 100 km, the fringe field becomes negligibly small.

Figure 4c shows results for configuration-2 having L=400 km, W=1200 km, and d=200 km. Interestingly, in this case, fringe field does not decrease as fast as that of configuration-2 having d=100 km both vertically and along north-south, implying that fringe field can easily penetrate downward and extend along north-south if the width is increased.

Next, we present how fringe field would vary in the east-west direction when more than one plasma bubble exists. Since more than one plasma bubble can occur in the east-west direction and the zonal dimension of bubble as well as inter-bubble separation could vary from one day to another, we have examined variation of fringe field along east-west in such configurations. For demonstration, we have considered a pair of bubble separated by a distance. We have chosen two cases, one having east-west bubble dimension of 100 km and inter-bubble separation of 300 km and in another the east-west bubble dimension is changed to 200 km keeping the inter-bubble separation same (300 km).

In order to perform computation for the first case, we have considered two parallel plate capacitors having d =100 km each and separated by 300 km, which is shown in Figure 5a. This configuration signifies that there are two plasma bubbles having width of 100 km each and inter-bubble separation of 300 km. Although in principle both width and inter-bubble separation could be different from those used here, this configuration is used to demonstrate how the fringe field would vary both in east-west and vertical directions. Figure 5b shows variations in fringe field in the east-west direction and at different altitudes. As can be noted that fringe field exists everywhere in the east-west direction, be it just below the capacitor or in the region where there is no capacitor. We, however, note that fringe field is maximum where the central part of the capacitor is located and minimum at the farthest distance from the centre of the capacitor. Again fringe field decreases with decreasing height. Figure 6a-b show results similar to that of Figure 5 but for d=200 km. In this case, fringe field does not even drop in between the two capacitors. It looks as if there is one capacitor. This implies that if two bubbles are wide enough and the inter-bubble separation is of the order of the bubble width, fringe field appears to be equivalent to that of a much wider bubble than a single bubble.

In summary, we have shown that (1) fringe field decreases with decreasing altitude, (2) fringe field decreases significantly as we move away from the magnetic equator, (3) fringe field at a given height and latitude increases with increasing bubble width, and (4) in case of multiple bubbles, large fringe field is observed at the magnetic equator in regions where there exists no overhead bubbles.

The finding that fringe field decreases with decreasing altitude is consistent with that noted earlier by Alam Kherani et al. [2004] and suggests that a highly depleted plasma bubble providing strong polarization charges would be required to produce necessary fringe field at the valley- and E-region for the upward transportation of plasma and plasma irregularities. But a more important finding relevant for understanding the occurrence (non-occurrence) of valley region irregularities at the equator (low latitudes) is the reduction of fringe field with latitude. It is also interesting to note that at low latitude the fringe field just below the bubble decreases at a much faster rate than that at lower altitudes. Given the fact that plasma bubble develops vertically at the equator and maps to higher latitudes along field lines simultaneously, the polarization process in the low latitude close to the magnetic equator is believed to be quite similar and the fringe fields both at the equator and low latitudes on a local consideration would be same. But the fact that bubbles are magnetic field-aligned and the fringe field at a low latitude point beneath the plasma bubble would be the vector sum of all field lines passing through it, the fringe fields are smaller than that at the equator. This reduction of fringe field is presumed to be one of the important reasons for impeding vertical transport of plasma and plasma irregularities at low latitudes.

We also found that the fringe field increases with increasing bubble width and in the case of multiple bubbles, fringe fields can be present even at locations where overhead bubble does not exist. This finding supports Jicamarca observations (shown in Figure 1) that valley region irregularities are observed in a region where there exists no overhead bubble [Woodman and Chau, 2001].

**5. Concluding remarks**

Results presented in this paper showed that observing valley region irregularities during the initial phase of equatorial plasma bubble at the equator and not observing at low latitudes could be linked with varying fringe field with latitude linked with the field-aligned nature of equatorial plasma bubble. Although it is not known whether fringe field can generate valley region irregularities, their presence and magnitude would be able to induce vertical transport of plasma or plasma irregularities and their variation from equator to low latitude at least partly is consistent with observations. Further work is necessary to understand their occurrence/non-occurrence and their linkage to the equatorial plasma bubble parameters, viz., bubble dimension, depletion level and height of occurrence.

**Acknowledgments.** The work was done at NARL when S Mukherjee was visiting as a Summer Fellow of the Indian Academy of Sciences. SM gratefully acknowledges the Indian Academy of Sciences for supporting his research at NARL.

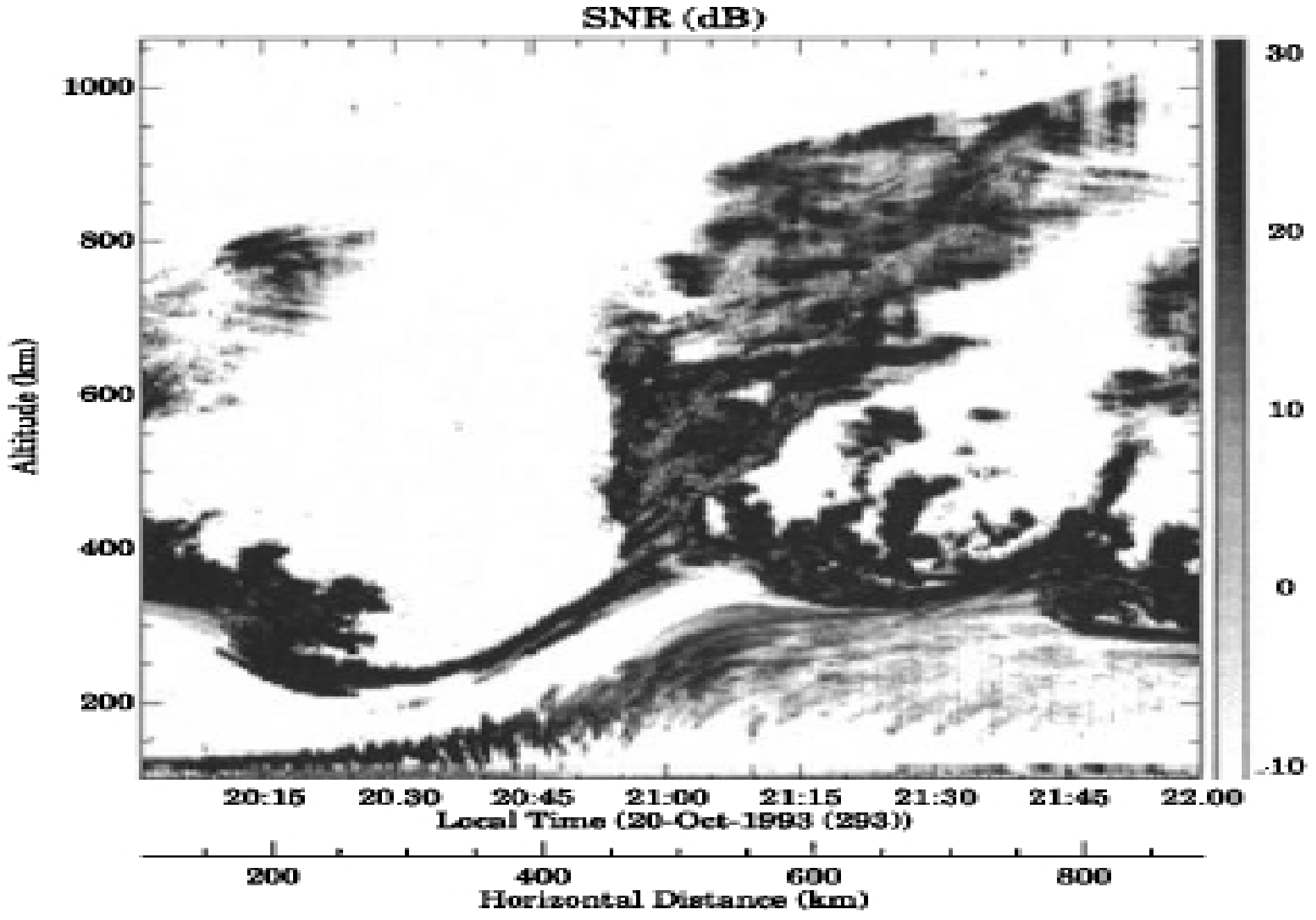

**Figure 1.** Height-time-SNR map of ionospheric irregularities observed in the E and F regions by the Jicamarca radar (reproduced from Woodman and Chau, 2001). Note the ascending echoing structures in the valley region during the updraft of the F region irregularities and plume structures.

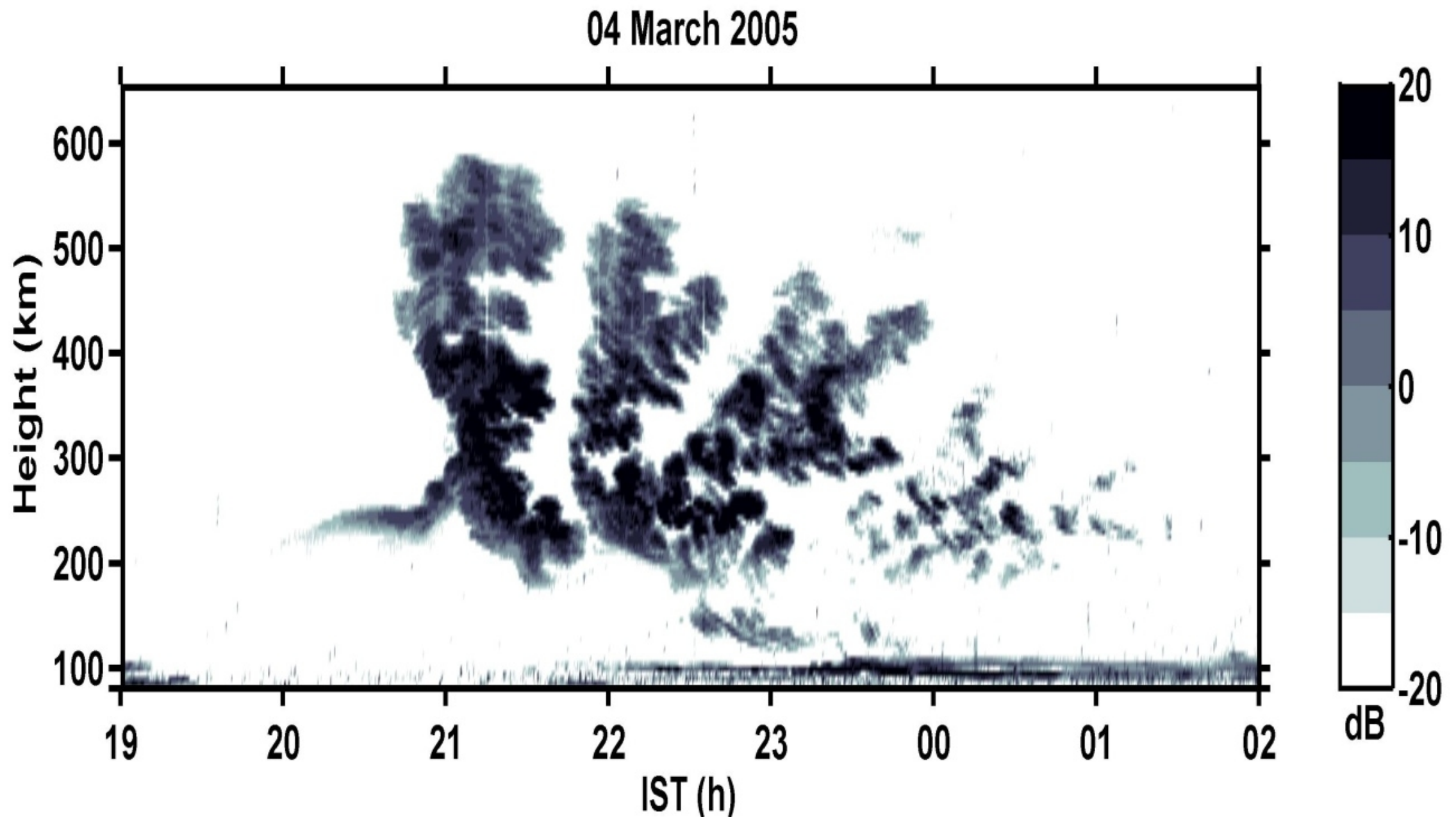

**Figure 2.** Height-time-SNR map of ionospheric irregularities observed by the Gadanki radar. Note that there are no valley region irregularities during the updraft of the F region irregularities and plume structures.

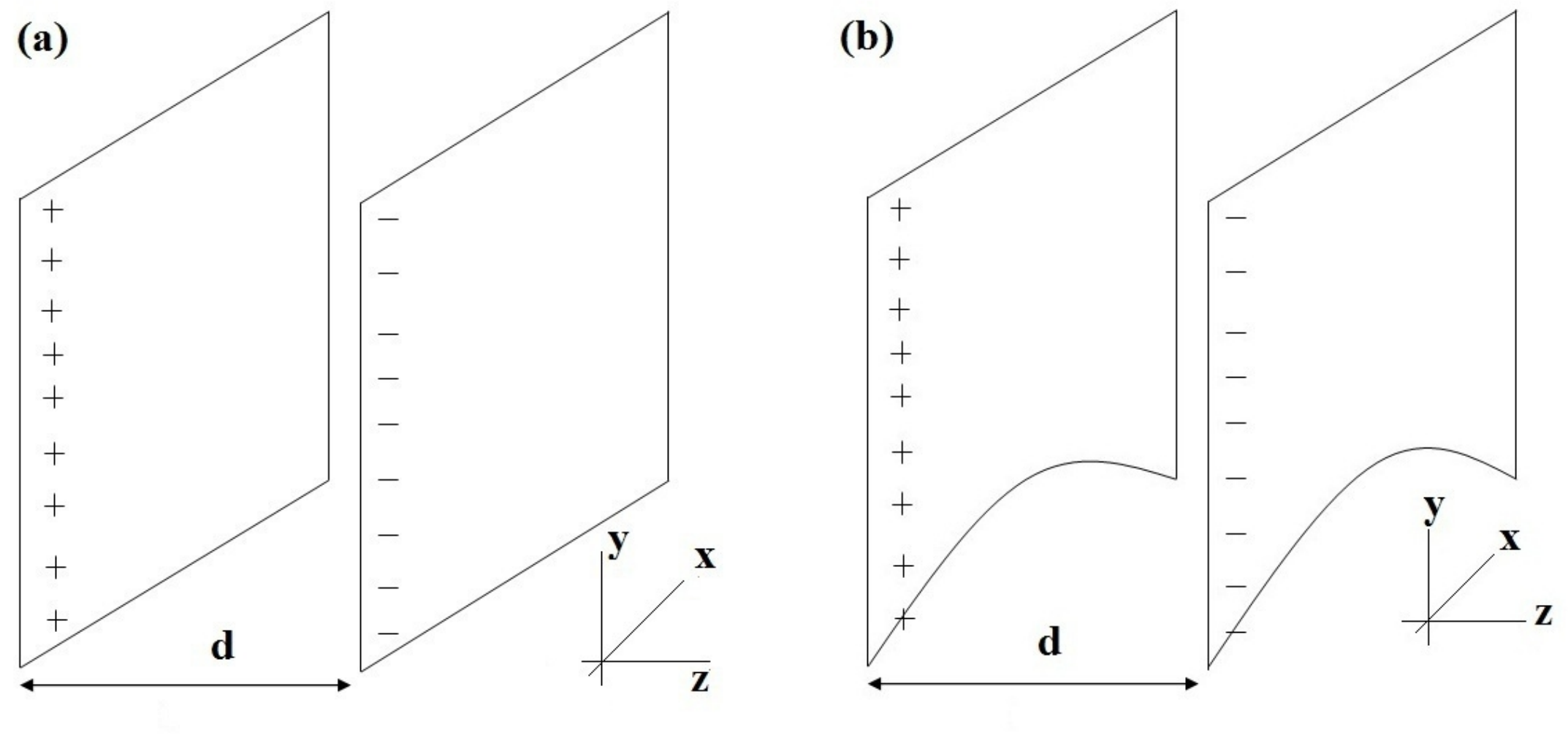

**Figure 3.** Parallel Plate Capacitors with (a) flat edge at the bottom and (b) curved edge at the bottom. The coordinate system is also shown.

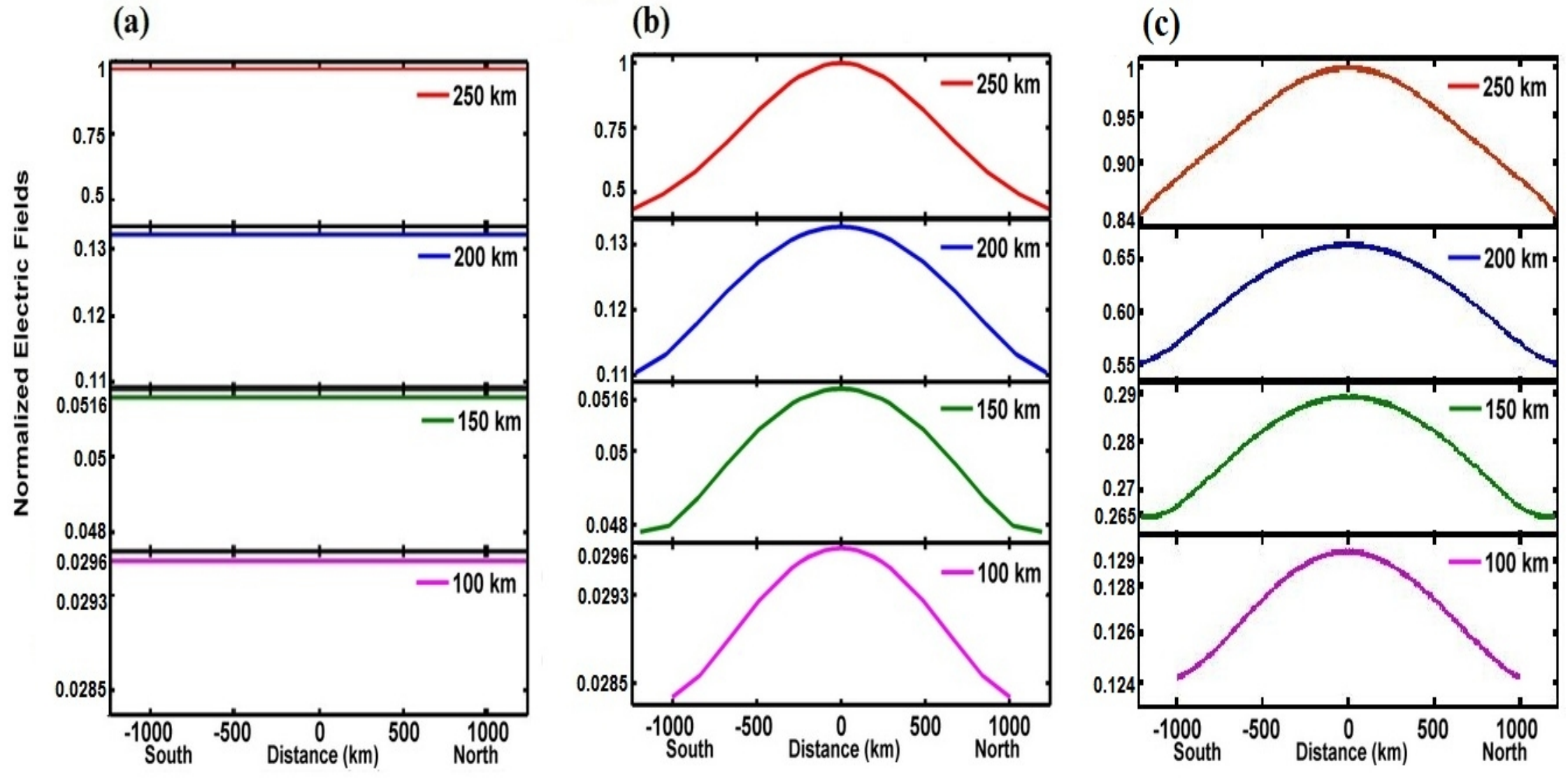

**Figure 4 (a).** Distributions of normalized electric field as a function of height and latitude along the North-South direction when the capacitor is having the configuration of Figure 3b. **(b).** Similar distribution when the capacitor is having the configuration of Figure 3a.

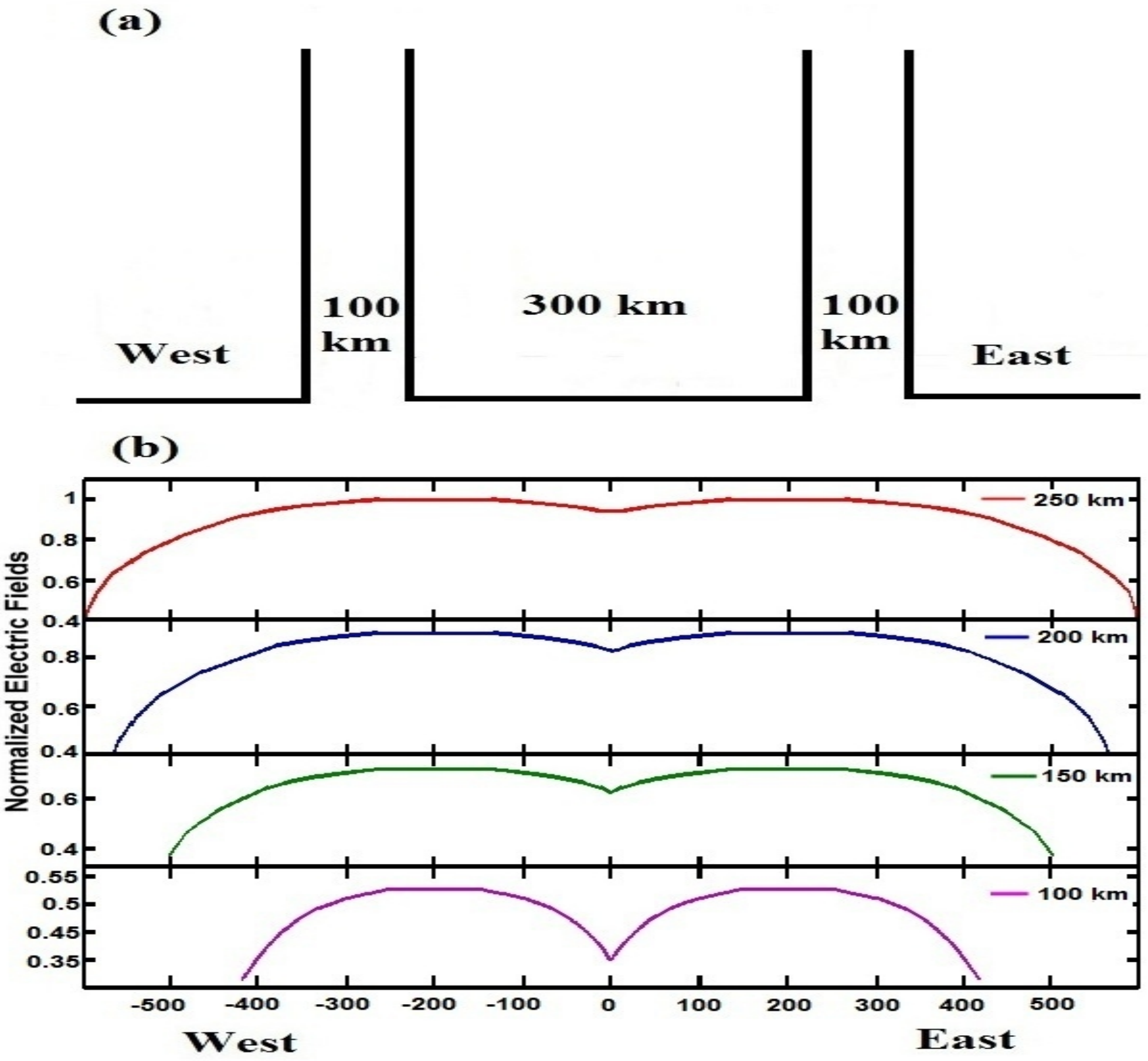

**Figure 5.**

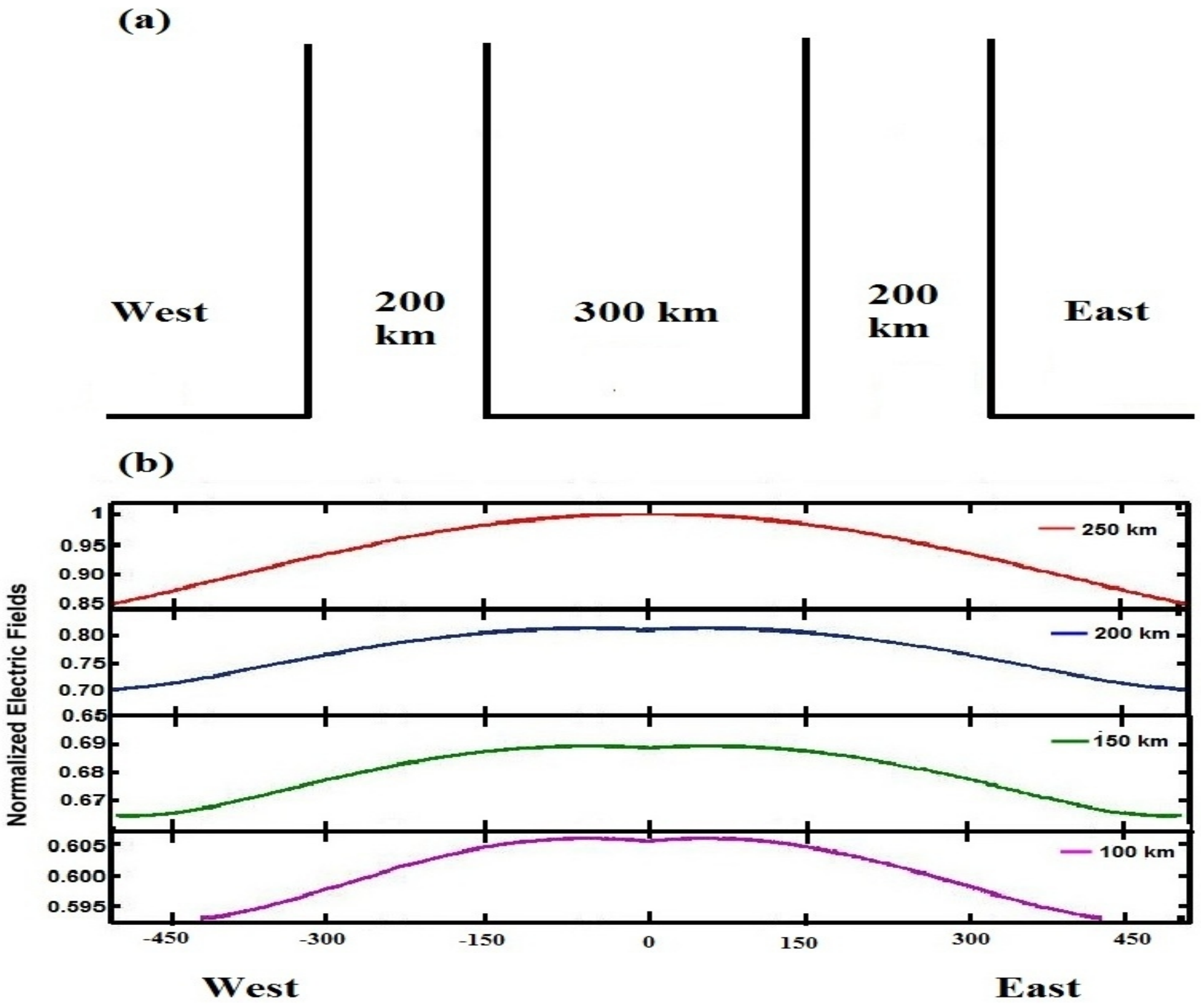

**Figure 6:**